\documentclass[aps,prb,twocolumn,showpacs,preprintnumbers,amsmath,amssymb,longbibliography]{revtex4-2}

\newcommand{\bra}[1]{\langle #1|}
\newcommand{\ket}[1]{|#1\rangle}

\newcommand{\be}{\begin{equation}}
\newcommand{\ee}{\end{equation}}
\newcommand{\bea}{\begin{eqnarray}}
\newcommand{\eea}{\end{eqnarray}}

\newcommand{\fig}[1]{Fig.~\ref{#1}}

\newcommand{\e}{\varepsilon}

\newcommand{\s}{\sigma}

\usepackage{float}
\usepackage[dvipsnames]{xcolor}
\usepackage{ulem}
\usepackage{verbatim} 
\usepackage{amsmath}
\usepackage{graphicx}
\usepackage{dcolumn}
\usepackage{bm}
\usepackage{hyperref}

\definecolor{oucrimsonred}{rgb}{0.6, 0.0, 0.0}

\begin{document}


\title{Nonmonotonic buildup of spin-singlet correlations in a double quantum dot}

\author{Kacper Wrze\'sniewski}
\email{wrzesniewski@amu.edu.pl}
\affiliation{Institute of Spintronics and Quantum Information,
	Faculty of Physics, 
	Adam Mickiewicz University,
	Uniwersytetu Pozna\'nskiego 2, 61-614 Pozna\'n, Poland}

\author{Tomasz \'Slusarski}
\affiliation{Institute of Spintronics and Quantum Information,
	Faculty of Physics, 
	Adam Mickiewicz University,
	Uniwersytetu Pozna\'nskiego 2, 61-614 Pozna\'n, Poland}

\author{Ireneusz Weymann}
\email{weymann@amu.edu.pl}
\affiliation{Institute of Spintronics and Quantum Information,
Faculty of Physics, 
Adam Mickiewicz University,
Uniwersytetu Pozna\'nskiego 2, 61-614 Pozna\'n, Poland}

\date{\today}


\begin{abstract}
Dynamical buildup of spin-singlet correlations between the two quantum dots
is investigated by means of the time-dependent numerical renormalization group method.
By calculating the time-evolution of the spin-spin expectation value upon
a quench in the hopping between the quantum dots,
we examine the time scales associated with the development
of an entangled spin-singlet state in the system.
Interestingly, we find that in short time scales the effective exchange interaction
between the quantum dots is of ferromagnetic type, favoring spin-triplet correlations,
as opposite to the long time limit, when strong antiferromagnetic correlations develop
and eventually an entangled spin-singlet state is formed between the dots.
We also numerically determine the relevant time scales and show
that the physics is generally governed by the interplay between the Kondo correlations
on each dot and exchange interaction between the spins of both quantum dots.
\end{abstract}

\maketitle

\section{\label{sec:level1}Introduction}

Double quantum dot systems coupled by a tunable tunnel barrier are important,
prototypical structures allowing for convenient control and manipulation
of the electronic occupation and spin degrees of freedom
\cite{vanderWiel2002Dec, Hayashi2003Nov, Gorman2005Aug, Obata2010Feb,
	Brunner2011Sep, Ramon2011Oct, Liu2012Jan,
	Nguyen2013Dec, Busser2013Dec, Busser2014Nov, Hao2014May, Hiltunen2015Feb, Mehl2015Jan, Busser2023Mar}.
Such systems have more complex electronic structure
than single quantum dots \cite{Goldhaber-Gordon1998Jan,Kouwenhoven2001Jun,Hanson2007Oct}
and therefore reveal further interesting properties and correlations effects
\cite{Potok2007Mar,Keller2014Feb,Keller2015Oct,Bordoloi2022Dec,Dvir2023Feb}.
Moreover, the ability to trap and control electron spins in quantum dots opens up 
new possibilities for promising applications in the field of quantum computing
\cite{Loss1998Jan, Burkard1999Jan, DiVincenzo2000Nov,
	Taylor2005Dec,DiVincenzo2005Sep,Stepanenko2007Feb, Fischer2009Sep, Kim2011Mar}.
In fact, the generation and manipulation of quantum states of individual electrons
in the dots enabled the creation of qubits with high fidelity and long coherence times
\cite{Petta2005Sep, Nowack2007Nov, Buscemi2011Jan, Kloeffel2013Mar, Medford2013Jul,
	Kawakami2014Sep, Kim2014Jul, Veldhorst2014Dec, Muhonen2014Dec,
	Vandersypen2017Sep, Nichol2017Jan, Qiao2020Jun}.
Coupled double quantum dot systems are in particular widely considered as hosts for exchange-qubits,
since they grant an overall tunability and increased stability due to noise and decoherence suppresion.
Additionally, such systems  provide several schemes for fast manipulation
and also the possibility to form three-electron qubits
\cite{Pierre2009Dec, Churchill2009Apr, Barthel2010Apr, Reynoso2011Nov,
	Shi2012Apr, Shi2014Jan, Knapp2016Mar, Cao2016Feb, Russ2017Aug}.

Nevertheless, for efficient exploitation of 
quantum dots in solid-state quantum information, it is 
of importance to fully understand the relevant dynamical behavior,
when also correlation effects may play an important role.
In this regard, the dynamics and transient behavior of double quantum dot systems
remains still a quite unexplored problem, demanding very precise theoretical
and experimental studies \cite{Coish2005Sep, Ramon2010Jan, Hu2011Apr,
	Khomitsky2012Mar, Hung2013Aug, Pal2014Jan, Ferraro2015Feb,
	Urdampilleta2015Aug, Maslova2017Oct, Maslova2018Feb, Mantsevich2019Aug, Lan2020May, Taranko2021Apr, Baran2021May}.
In this work we therefore focus on theoretical study of quench dynamics
of a half-filled double quantum dot 
following an abrupt switching on of the coupling between the two quantum dots,
that were initially coupled to external electronic reservoirs, but isolated from each other.
The dynamical buildup of spin-singlet correlations between the two quantum dots
is investigated by means of the time-dependent numerical renormalization
group (tdNRG) method \cite{Anders2005Oct, Bulla2008Apr, Nghiem2014Jul,
	Nghiem2014Feb, Nghiem2018Oct, Wrzesniewski2019Jul}.
By calculating the time-evolution of the spin-spin expectation value upon a quench,
we examine the time scales associated with the development
of an entangled spin-singlet state in the system.
We analyze the development of the spin correlations
between the dots mediated by a direct hopping between the dots.
We show that the spin-singlet correlations start to compete with Kondo correlations
formed in the initial state, and for considerable values of exchange
interaction lead to a singlet state formed between two dots.
Interestingly, when the dynamics is driven by exchange coupling
values significantly exceeding the Kondo temperature $T_K$,
there is a development of ferromagnetic order at short time scales,
which precedes the formation of an entangled spin-singlet state in the long-time limit.

The following is an outline of how this paper is organized. In Sec. \ref{section:Model}
we describe the model and briefly introduce the method used in numerical calculations.
In Sec. \ref{section:Static} the relevant static properties of the system
are presented and discussed, while the main results and discussion
of spin dynamics are in Sec. \ref{section:Dynamics}.
The paper is concluded in Sec. \ref{section:Conclusions}.

\section{Model and method} \label{section:Model}

\begin{figure}[t]
	\includegraphics[width=0.9\columnwidth]{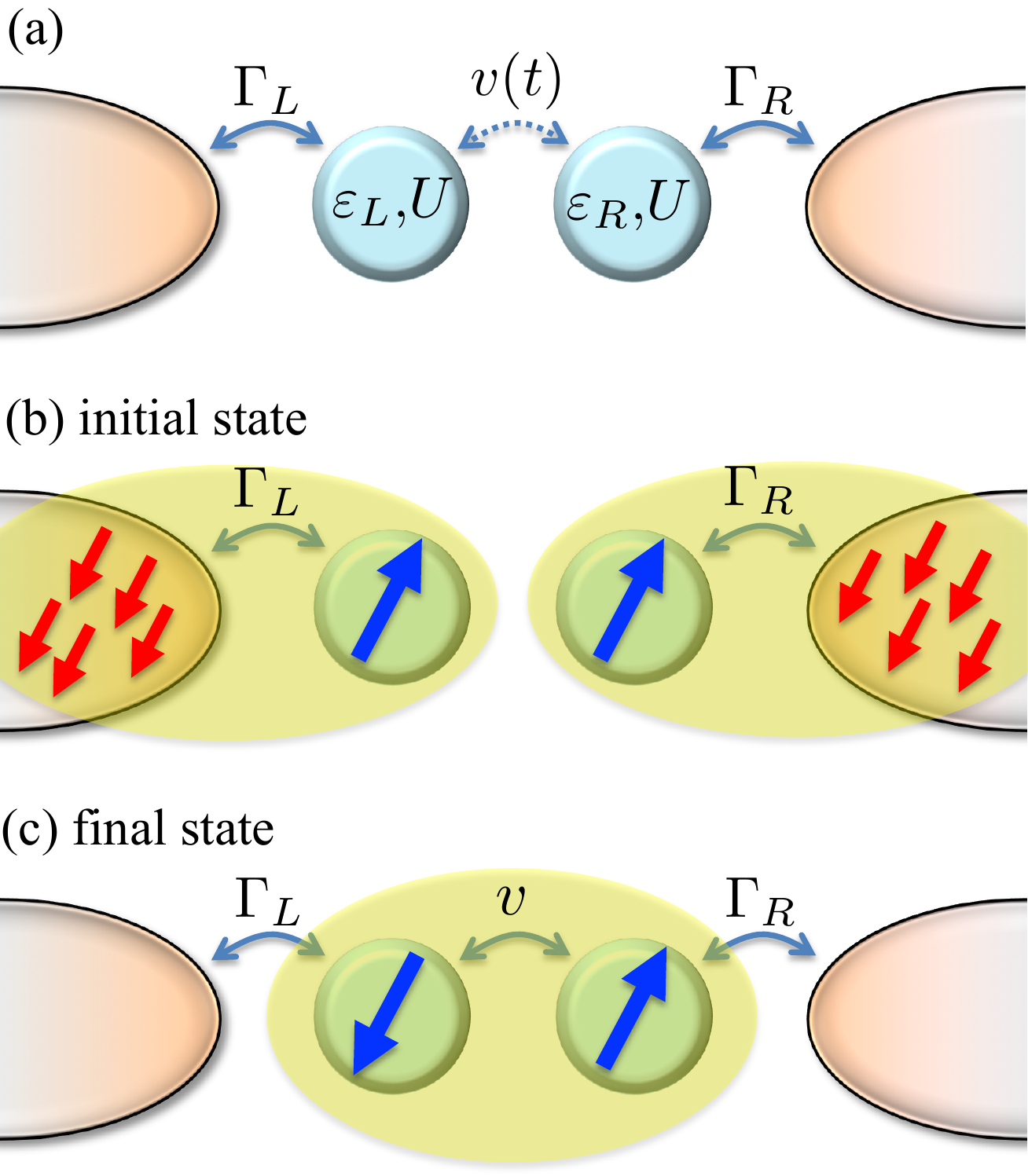}
	\caption{(a) The schematic of the consider double quantum dot system.
		Each quantum dot is coupled to its metallic lead
		with coupling strength $\Gamma_\alpha$,
		with $\alpha=L$ ($\alpha=R$) for the left (right) lead.
		$\e_L$ and $\e_R$ denote the orbital level energies in the corresponding dots,
		while $U$ stands for the Coulomb correlations.
		The quantum dots are coupled through the hopping matrix element $v(t)$. 
		We study the evolution of the system following the quench in $v(t)$,
		as schematically presented in (b) and (c).
		(b) In the initial state, $t<0$, $v=0$ and Kondo state
		develops separately on each quantum dot.
		(c) In the final state, $t\to \infty$, singlet-state
		forms between the two quantum dots,
		provided that the antiferromagnetic exchange interaction between the dots
		is larger than the corresponding Kondo temperature.
		}
	\label{fig:schem}
\end{figure}

The considered system is presented in \fig{fig:schem}(a).
It consists of two quantum dots, each attached to its own metallic
reservoir, and coupled to each other through the hopping matrix element $v(t)$.
The Hamiltonian of such device can be written as
\begin{equation}
	H = H_{r} + H_{DD} + H_{T},
\end{equation}
where the first term describes the electrodes in the free quasi-particle approximation
\begin{equation}
   H_{r}= \sum_{\alpha=L,R} \sum_{k\s} \e_{\alpha k\s} c^\dagger_{\alpha k\s} c_{\alpha k \s}.
\end{equation}
Here, $c^\dagger_{\alpha k\s}$ ($c_{\alpha k \s}$) creates (annihilates)
an electron of momentum $k$, spin $\s$ in the left ($\alpha=L$)
or right ($\alpha=R$) lead with the corresponding energy $\e_{\alpha k\s}$.
The second term $H_{DD}$ models the double quantum dot system
and is given by
\begin{eqnarray}
	H_{DD}&=&\! \sum_{\alpha=L,R} \sum_\s \e_{\alpha} d^\dagger_{\alpha \s} d_{\alpha  \s}
	+\! \sum_{\alpha=L,R}  U d_{\alpha\uparrow}^\dag d_{\alpha\uparrow} d_{\alpha\downarrow}^\dag d_{\alpha\downarrow} \nonumber\\
&+& v \sum_\s (d_{L\sigma}^\dag d_{R\s} + d_{R\sigma}^\dag d_{L\s}),
\end{eqnarray}
with $d^\dagger_{\alpha \s}$ ($d_{\alpha \s}$) being
the creation (annihilation) operator on the left ($\alpha=L$)
or right ($\alpha=R$) quantum dot for an electron of spin $\s$,
$\e_\alpha$ denotes the corresponding orbital level energy
and $U$ stands for the Coulomb correlations assumed to be equal for both dots.
The hopping amplitude between the two dots is given by $v$.
Without loss of generality, we assume that each dot is at half filling, $\e_\alpha = -U/2$,
and that the system is symmetric, $\Gamma_L = \Gamma_R = \Gamma$.

We are interested in the development of spin correlations between the two
quantum dots when turning on the hopping $v$, see Figs.~\ref{fig:schem}(b)-(c).
In particular, we determine the time dependence of the spin-spin expectation value,
\begin{equation}
	S_{LR}(t) \equiv \bra{\Psi(t)} \vec S_L\cdot \vec S_R \ket{\Psi(t)},
\end{equation}
which quantifies the formation of the singlet state between the quantum dots.
Here, ${\vec S_\alpha = \tfrac{1}{2}\sum_{\s\s'} d^\dag_{\alpha \s} \vec{\s}_{\s\s'}d_{\alpha \s'}}$
is the spin operator of the dot $\alpha$, with $\vec{\s}$ denoting the vector of Pauli spin matrices.
The many-body state of the system $\ket{\Psi(t)}$ evolves according to the full Hamiltonian $H$
\begin{equation}
	\ket{\Psi(t)} = e^{-iHt}\ket{\Psi(0)},
\end{equation}
where $\ket{\Psi(0)}$ is the initial state that we prepare assuming $v=0$.
To be able to resolve the system's dynamics in the
presence of strong electronic correlations in the most accurate manner,
we employ here the time-dependent numerical renormalization
group method \cite{Anders2005Oct, Bulla2008Apr, Nghiem2014Jul, Nghiem2014Feb, Nghiem2018Oct, Wrzesniewski2019Jul}.

The general form of the considered time-dependent Hamiltonian is given by
\begin{equation}
	H(t) = \Theta(-t)H_0 + \Theta(t)H.
\end{equation}
Here, the Hamiltonian $H_0$ describes the system before quantum quench in initial equilibrium  for $t<0$.
Then, the final Hamiltonian $H$ models the system for $t\geq 0$, when a sudden change in
$H$ is performed with respect to $H_0$.
In the considered case, it is switching on the hopping between
the quantum dots in a step-like fashion at time $t=0$.
In tdNRG both Hamiltonians are diagonalized in the Wilson chain geometry by
using the numerical renormalization group method
\cite{Wilson1975Oct, Weichselbaum2007Aug, Bulla2008Apr, Andreas2012, FlexibleDMNRG}.
The discarded states determined in this iterative procedure, including all states found in the last iteration,
are used to build a full many-body eigenbases of the Hamiltonians.
In the next step, we determine the dynamical quantities.
In order to obtain the relevant time dependencies, the discrete data in form of Dirac delta peaks
is collected and then Fourier transformed into the time domain
\cite{Andreas2012,Wrzesniewski2019Jul}.
It is important to mention that tdNRG is a fully nonpertirbative method,
and as such, it allows us to obtain high quality data
where all correlations, also the ones driving the Kondo effect,
are taken into account on an equal footing.

\section{Static properties} \label{section:Static}

\begin{figure}[t]
	\includegraphics[width=0.99\columnwidth]{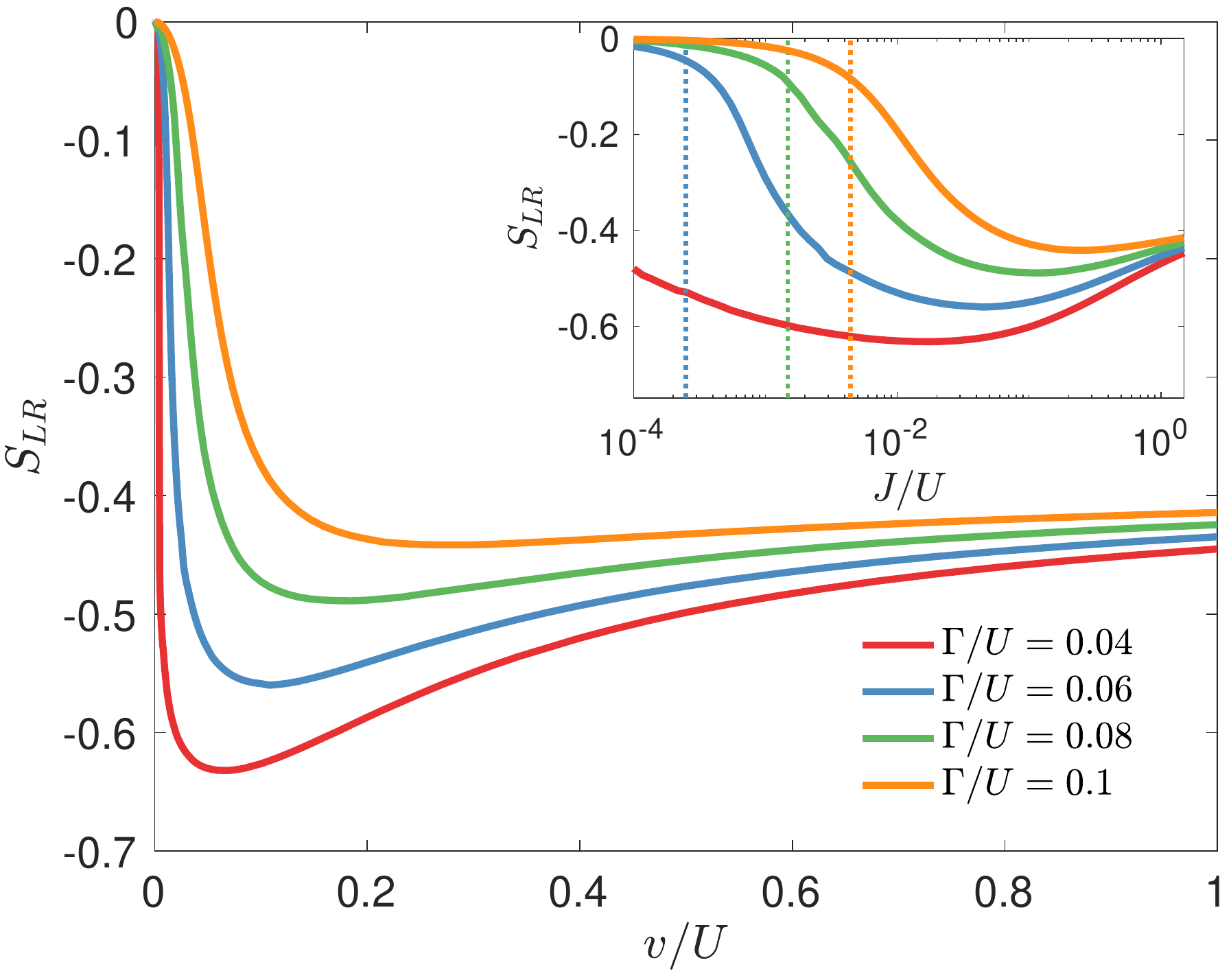}
	\caption{The static expectation value of $S_{LR}$
		calculated as a function of the hopping between the quantum dots $v$
		for different values of the coupling to the contacts $\Gamma$, as indicated.
		The inset presents the dependence of $S_{LR}$
		on the exchange interaction $J$,
		where the dashed lines indicate the Kondo temperature $T_K$
		for corresponding values of $\Gamma$,
		as estimated from the Haldane's formula \cite{Haldane1978}.
		Note the logarithmic x-axis scale in the inset.
		The parameters are: $\e_{L}=\e_{R}=-U/2$ and $U=0.1$
		in units of band halfwidth.
	}
	\label{Fig:S1S2static}
\end{figure}

To begin our discussion, in \fig{Fig:S1S2static} we present the static expectation value of 
$S_{LR}$ calculated as a function of the hopping between the dots
for different values of the coupling to external contacts. One can see that 
$S_{LR}$ depends in a nonmonotonic manner on $v$.
To understand this behavior, one needs to realize that there are two competing
energy scales in the system. The first one is associated with the Kondo effect,
which, in the absence of hopping between the dots, develops in each quantum dot separately.
The Kondo temperature depends in an exponential way on the ratio of $U/\Gamma$
as predicted by the Haldane's formula \cite{Haldane1978}, 
$T_K = \sqrt{U\Gamma/2}\; {\rm exp}(-\pi U/8\Gamma)$.
On the other hand, the hopping $v$ generates an antiferromagnetic exchange interaction
between the dots, which can be estimated from the singlet-triplet splitting
\begin{equation}
	J = \frac{1}{2}\left[\sqrt{16 v^2 + U^2} -U\right].
\end{equation}
For small values of $v$ it can be approximated by $J\approx 4v^2/U$.
When each quantum dot is singly occupied, as assumed in our considerations,
the physics is governed by the ratio of $J/T_K$, as can be seen in the inset
of \fig{Fig:S1S2static}, which presents $S_{LR}$ plotted as a function of $J$ on a logarithmic scale,
while the vertical dotted lines indicate the corresponding Kondo temperatures.
Clearly, the correlations $S_{LR}$ become considerable when 
$J\gtrsim T_K$. Then, at some even larger value of $J$, $S_{LR}$ exhibits
a local minimum, with $S_{LR} \to -3/4$ for $\Gamma \ll U$,
to raise again with further increase of $J$.
This further increase (decrease in magnitude)
can be explained by considering all the double
quantum dot states, the energies of which strongly depend on $v$,
such that for larger $v$, other states become populated and 
$S_{LR}$ increases, see \fig{Fig:S1S2static}.

\section{Buildup of spin-singlet correlations} \label{section:Dynamics}

\begin{figure}[t]
	\includegraphics[width=0.99\columnwidth]{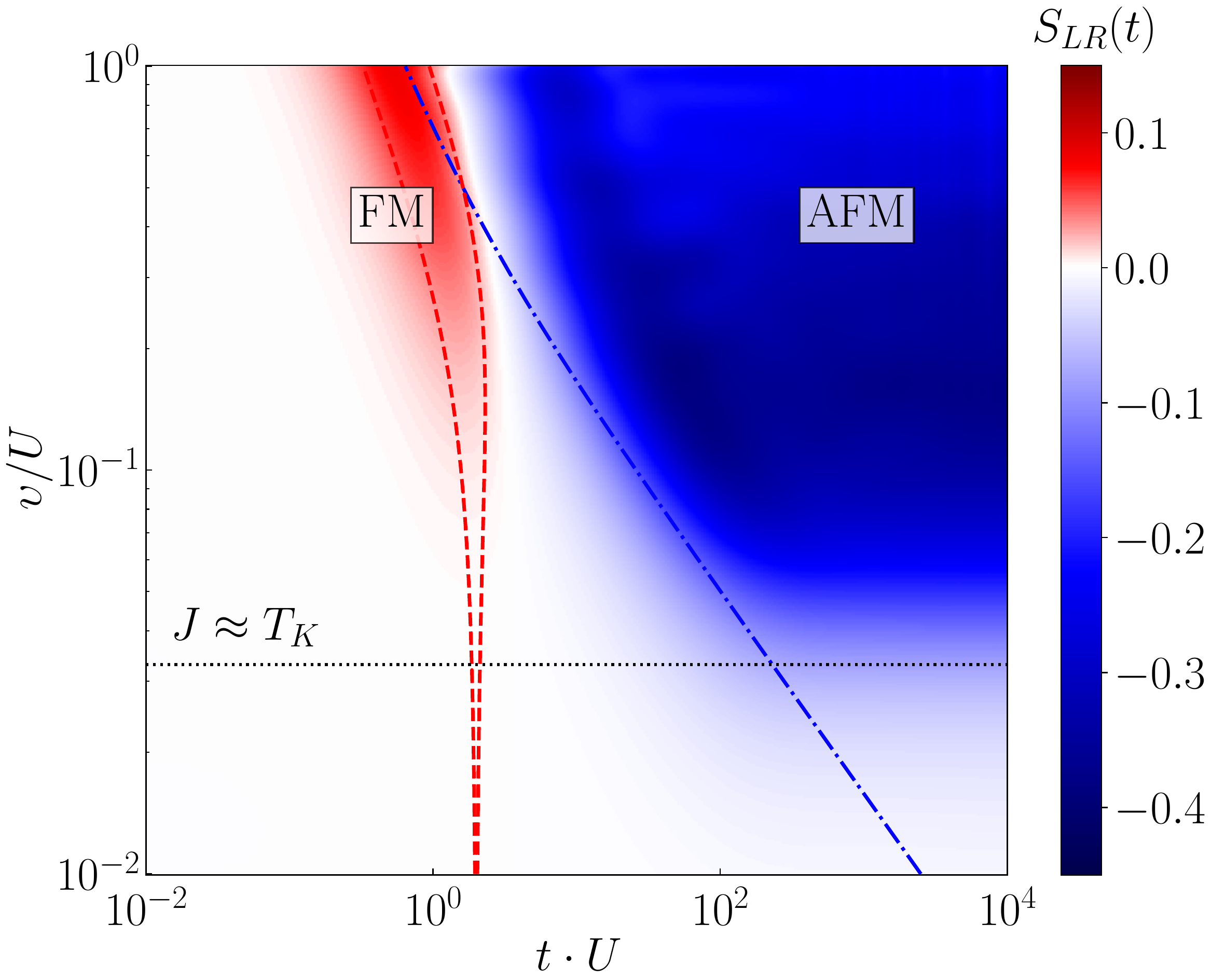}
	\caption{The expectation value $S_{LR}(t)$
		calculated as a function of time $t$ and the hopping between the dots $v$.
		The system is prepared in the initial state with $v=0$
		and then time-evolved according to the full Hamiltonian
		with a finite value of $v$. The dotted line presents the
		value of $v$ for which $J\approx T_K$, see the main text for details.
		In short time scales the two spins on the dots exhibit
		temporary ferromagnetic (FM) order, $S_{LR}(t)>0$, while 
		for longer times the system shows considerable spin-singlet correlations,
		i.e. antiferromagnetic (AFM) order with $S_{LR}(t)<0$.
		The dotted-dashed (dashed) line presents the time scale
		associated with excitation to the triplet (doublet) state.
		The other parameters are the same as in \fig{Fig:S1S2static}.
	}
	\label{Fig:S1S2dyn2D}
\end{figure}

Let us now analyze the dynamical behavior of the spin-spin expectation value $S_{LR}(t)$,
following the quench in the hopping between the dots.
In the initial state the system consists of two identical, disconnected copies
of quantum dot attached to metallic electrode, each displaying the Kondo effect,
see Fig.~\ref{fig:schem}(b).
Then, we turn on the hopping between the dots and study
how the entanglement builds up in the system, quantified by the spin-spin
expectation value. The time evolution of $S_{LR}(t)$ for different values
of the hopping between the dots is shown in \fig{Fig:S1S2dyn2D}.
First of all, one can note that finite value of $S_{LR}(t)$ develops
only when the hopping $v$ is sufficiently large. The dotted line
in the figure indicates the value of $v$ for which the effective exchange interaction $J$
becomes of the order of the Kondo temperature, $J\approx T_K$.
For assumed parameters, this happens when $v/U\approx 0.033$.
Indeed, when $J<T_K$, the Kondo singlet state formed on each dot
dominates and the entangled singlet state between the two dots
hardly develops.
However, once $J \gtrsim T_K$, see $v/U\gtrsim 0.033$ in the figure,
$S_{LR}(t)$ exhibits considerable values in the long time limit,
indicating spin-singlet (antiferromagnetic) correlations between the quantum dots.
This is schematically displayed in Fig.~\ref{fig:schem}(c).
For intermediate values of $v$,
the time scale at which the AFM correlations develop
can be related to the singlet-triplet excitation energy,
$t\sim 1/J$, see the dotted-dashed line in \fig{Fig:S1S2dyn2D}.

Interestingly, in the intermediate time regime, $t\approx 1/U$, 
we observe a region when the spin-spin expectation value is positive.
This implies ferromagnetic correlations between the quantum dots.
In other words, before the singlet state forms in the double dots, the calculated system dynamics
reveals that in short time scales the effective exchange interaction changes sign,
favoring triplet alignment of quantum dot spins. 
The boundary of this dynamical regime,
indicated by the red dashed lines in \fig{Fig:S1S2dyn2D}, is established
by the two time scales corresponding to the excitations to the doublet states.
Note that the energy of doublet states splits with increasing $v$,
encompassing the region where $S_{LR}(t)>0$, see \fig{Fig:S1S2dyn2D}.
If one considers realistic quantum dot parameters,
e.g. $U\sim 1$ meV, this would result in time scale on the order of $t\sim 0.01-0.1$ ns.
As far as the mechanism of this sign change is concerned,
by comparing with the corresponding calculations for the two impurity Kondo model (not shown),
we can infer that it is due to the charge fluctuations to the doublet states
(such sign change is absent in the Kondo model).
As the system evolves toward the long time limit, the time scale associated with
the excitation to the triplet state (indicated by the blue dotted-dashed line)
is surpassed and followed by the build-up of spin-singlet correlations.

\begin{figure}[t]
	\includegraphics[width=0.99\columnwidth]{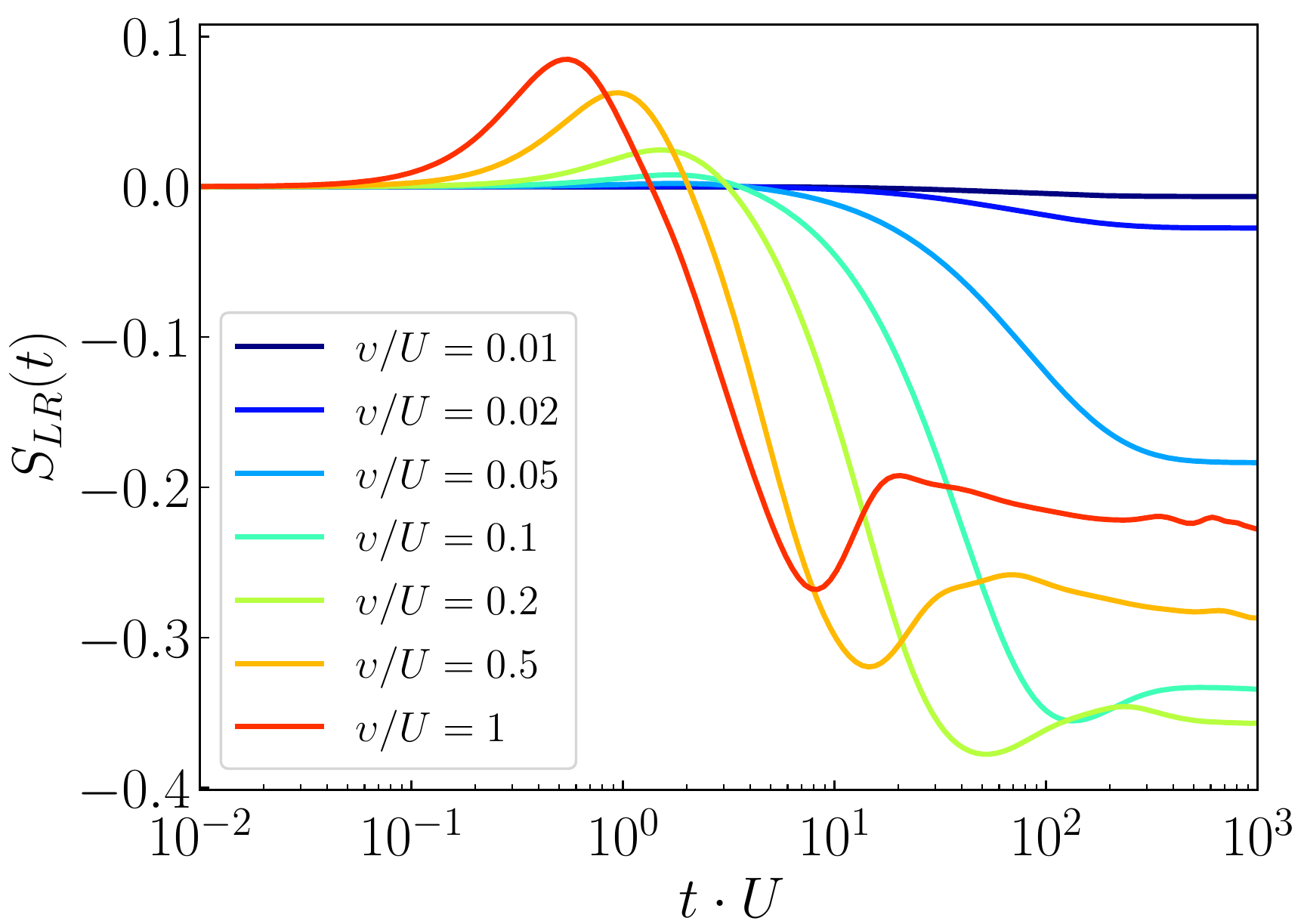}
	\caption{The time-evolution of the expectation value $S_{LR}(t)$
		calculated as a function of time for selected values of the hopping between the dots.
		The other parameters are the same as in \fig{Fig:S1S2static}.
	}
	\label{Fig:S1S2dyn}
\end{figure}

The detailed behavior of $S_{LR}(t)$ is shown in \fig{Fig:S1S2dyn},
which presents the cross-sections of \fig{Fig:S1S2dyn2D}
for selected values of the hopping $v$, as indicated.
One can now explicitly see that for small hoppings,
such that $J\lesssim T_K$, $S_{LR}(t)$ is generally suppressed
with a relatively low value in the long time limit.
However, when $J\gtrsim T_K$, see the case of $v/U\gtrsim 0.05 $ in 
\fig{Fig:S1S2dyn}, the absolute value in the long-time limit increases.
Moreover, the time scale when considerable antiferromagnetic correlations
develop decreases as $v$ raises. As already mentioned, in the short time scale,
temporary ferromagnetic correlations develop between the quantum dots.
These are present when $v\gtrsim \Gamma$, see \fig{Fig:S1S2dyn}.

\section{Conclusions} \label{section:Conclusions}

In this work we have studied the buildup of correlations
between the two quantum dots quantified by the spin-spin expectation value $S_{LR}(t)$.
By employing the numerical renormalization group in time domain,
we were able to include all correlations effects in a fully nonperturbative manner.
In particular, we have examined the time evolution of $S_{LR}(t)$
upon turning on the hopping between the two quantum dots.
It has been found that at small time scales, $t\sim 1/U$, $S_{LR}(t)$ becomes first positive,
indicating ferromagnetic exchange interactions, while only for larger times
the effective exchange interaction becomes antiferromagnetic
and an entangled spin singlet state is formed between the dots. 
The formation of such singlet state is conditioned by the value of hopping
between the quantum dots; it develops for such $v$ that the effective exchange interaction
$J\approx 4v^2/U$ becomes larger than the corresponding Kondo temperature 
for each quantum dot. Then, in the course of time evolution,
the singlet state forms between the dots, winning over the two separate
Kondo singlet states. The time scale when this happens is approximately given by $1/J$,
i.e. the strength of exchange interaction between the dots.


\begin{acknowledgments}
	This work was supported by the Polish National Science
	Centre from funds awarded through the decision Nos. 2017/27/B/ST3/00621
	and 2022/45/B/ST3/02826.
	We also acknowledge the computing time
	at the Pozna\'{n} Supercomputing and Networking Center.
\end{acknowledgments}

\bibliography{quench_dynamics_DQD}

\end{document}